\documentclass[11pt]{article}
\usepackage{graphicx}
\usepackage{cs12,html,hyperref}

\begin{document}
 
\title{A Systematic Spectroscopic X-Ray Study of Stellar 
      Coronae with XMM-Newton: Early Results}
 
\author{
M. G\"udel\altaffilmark{1}, 
M. Audard\altaffilmark{1}, 
K.~W. Smith\altaffilmark{1,2},
A. Sres\altaffilmark{1,2},
C. Escoda\altaffilmark{1},
R. Wehrli\altaffilmark{1,3},
E.~F. Guinan\altaffilmark{4},
I. Ribas\altaffilmark{4},
A.~J. Beasley\altaffilmark{5},
R. Mewe\altaffilmark{6},
A.~J. Raassen\altaffilmark{6},
E. Behar\altaffilmark{7},
H. Magee\altaffilmark{8}
}
\altaffiltext{1}{Paul Scherrer Institut, Switzerland}
\altaffiltext{2}{Institute of Astronomy, ETH Z\"urich, Switzerland}
\altaffiltext{3}{Alte Kantonsschule Aarau, Switzerland}
\altaffiltext{4}{Villanova University, PA, USA}
\altaffiltext{5}{Caltech, Pasadena, USA}
\altaffiltext{6}{SRON, The Netherlands}
\altaffiltext{7}{Columbia University, New York, USA}
\altaffiltext{8}{MSSL, UK}

\begin{abstract}
We have been conducting a comprehensive survey of stellar coronae with
the {\it XMM-Newton} Reflection Grating Spectrometers  during the 
commissioning, calibration, verification, and guaranteed time phases of
the mission,  accompanied by simultaneous observations with the EPIC cameras 
and, for several targets, with the radio VLA and/or the VLBA.
The principal aim of this project is threefold: i) To understand stellar
coronal structure and composition by studying systematics in the coronae
of stars with widely different levels of magnetic activity; ii) to investigate
heating and particle acceleration physics during flares, their role
in the overall coronal energy budget, and their possible role in the
quiescent stellar emission; iii) to probe stellar coronal evolution by studying
solar analogs of different ages. We report early results from this project.
\end{abstract}

\section{Introduction}
\begin{figure}[t!]
\vskip -0.3truecm
\hspace{1.0cm}
\includegraphics*[angle=0, width=11cm]{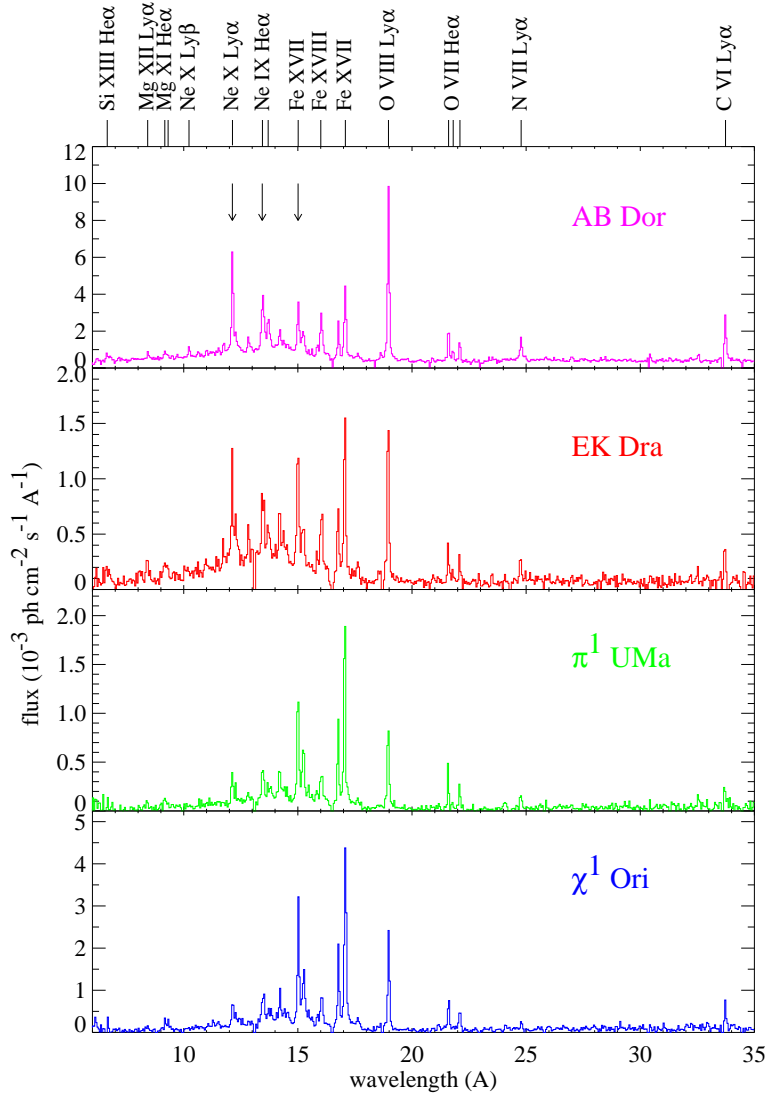}
\vskip -0.6truecm
\caption{Fluxed RGS X-ray spectra of four solar analogs. The overall coronal
activity decreases from top to bottom, while age increases. The arrows point
at \ion{Ne}{10}, \ion{Ne}{9}, and \ion{Fe}{17} lines that have similar maximum
line formation temperatures.}
\end{figure}

With the advent of {\it XMM-Newton} and {\it Chandra}, it has become possible
\index{XMM-Newton}
\index{X-ray spectroscopy}
for the first time to routinely investigate high-resolution stellar X-ray 
spectra. The {\it XMM-Newton} Reflection Grating Spectrometer (RGS) team
is conducting a comprehensive survey of coronae of cool stars, in particular 
to investigate i) stellar structure and composition, ii) coronal energy
release and heating, and iii) long-term evolution of stellar coronae. Complemented
by targets from the calibration and performance verification phase and some
guest observer targets, the survey comprises  26 targets. We present here selected
and preliminary results related to various issues.

\begin{figure}[t!]
\vskip -0.3truecm
\hspace{1.0cm}
\includegraphics*[angle=0, width=11cm]{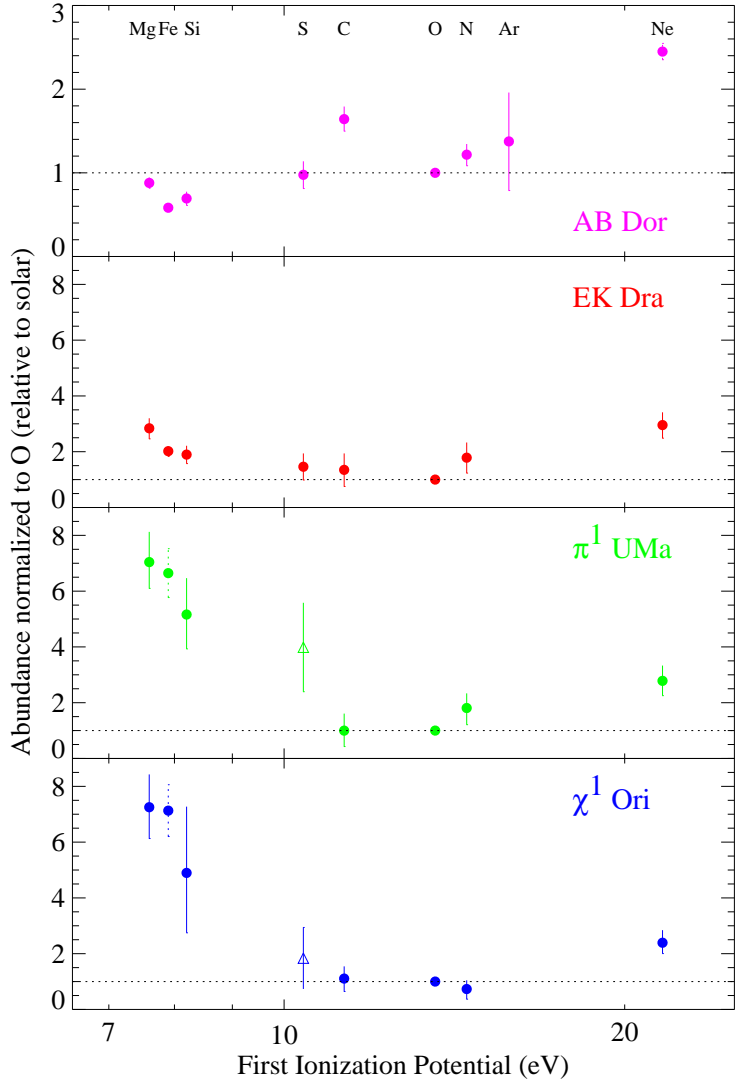}
\vskip -0.6truecm
\caption{Coronal elemental abundances of the same four solar analogs, relative
to the oxygen abundance and normalized with the respective solar photospheric 
ratios, as a function of FIP. Note the change from an IFIP effect (top) to a normal 
FIP effect (bottom) in the course of stellar evolution.}
\end{figure}

\section{Coronal Abundances}

\index{abundances, coronal}
\index{First ionization potential}
\index{FIP effect}
\index{IFIP effect}
Brinkman et al. (2001) reported an {\it inverse First Ionization Potential} (IFIP) effect in
the active corona of HR~1099, i.e., an increase of the elemental abundances with
\index{*HR~1099}
increasing FIP of the element, relative to solar photospheric abundances. Similar 
results were found for other very active stars (G\"udel et al. 2001ab) but not for less
active objects like Capella (Audard et al. 2001a). One of the fundamental problems 
\index{*Capella}
relates to the mostly unknown {\it photospheric} abundances of the respective stars 
(Audard et al. 2001b).
Since the coronal material ultimately derives from the photosphere, a physically
interesting result should relate coronal to {\it stellar} photospheric abundances.

We have studied a rather select sample of targets, namely young solar analogs with 
known photospheric metallicities - all of them are indistinguishable from 
solar photospheric. The stars vary essentially by their rotation periods (from 0.5 to 5 days)
and therefore activity levels (from $L_{\mathrm{X}} \approx 10^{30}$ to $\approx 10^{29}$~erg~s$^{-1}$).
The combined RGS1\&2 spectra are shown in Figure 1. Comparing the most active (AB Dor - with
a slightly later spectral type of K0~V)
with the least active ($\pi^1$ UMa and $\chi^1$~Ori) targets, one sees that
\index{*AB Dor}
\index{*EK Dra}
\index{*$\pi^1$ UMa}
\index{*$\chi^1$ Ori}
the continuum level is much lower in the latter, and the \ion{O}{8}/\ion{O}{7} 
flux ratios are smaller, indicating a lower coronal temperature for lower
activity. However, the maximum line formation  temperatures of the \ion{Ne}{10} and 
the \ion{Ne}{9} lines narrowly bracket the formation temperature of  \ion{Fe}{17},
the latter two ions being formed at very similar temperatures. These lines 
are marked by arrows  in Figure 1. It is evident, however, that the line ratios grossly
change from the most active to the least active star, indicating that the relative
{\it Fe abundance increases}. Figure 2 shows the abundance ratios of different
elements to oxygen from 
a full spectral analysis (in XSPEC), normalized to the photospheric ratios. We find clear evidence
for the inverse FIP effect in AB Dor, but a continuous change to a normal, solar-like
FIP effect in  the less active stars.

What is the cause for this transition? It is interesting to note that the non-thermal
coronal radio emission, attributed to gyrosynchrotron radiation from
\index{radio emission}
accelerated, high-energy electrons, rapidly drops from AB Dor to $\pi^1$ UMa, by a 
factor of at least 300 (Lim et al. 1992, Gaidos, G\"udel, \& Blake 2000).
One suspicion is that high-energy electrons penetrating into the stellar chromosphere
build up a (downward-pointing) electric field that pulls the chromospheric ions
(mostly from low-FIP elements) downward, thus preventing them  from reaching the 
corona, leaving a high-FIP enriched upper layer in the chromosphere that can access the
corona. As the non-thermal particle density 
decreases, this effect becomes less important, and a solar-like FIP effect can develop. For this
mechanism to work, a number of delicate conditions must be fulfilled. Most importantly,
the downward electron flux must not exceed the limit above which explosive chromospheric
evaporation develops as the energy cannot be radiated away sufficiently rapidly.
If evaporation does occur (e.g., during a flare) a larger part of the chromosphere will be
transported into the corona and the material will be mixed, so that a near-photospheric 
composition is recovered. This quenching of the IFIP effect has in fact been suggested 
from observations of large X-ray flares (e.g., G\"udel et al. 1999, Audard, G\"udel, \& Mewe
2001c).

\section{The Neupert Effect in Stellar Flares - Evidence for Chromospheric Evaporation?}

\index{Neupert effect}
\index{radio emission}
A standard flare scenario devised from many solar observations proposes
that accelerated coronal electrons precipitate
into the chromosphere where they lose their kinetic energy by collisions, thereby
heating the cool plasma to coronal flare temperatures. The subsequent overpressure
drives the hot material into the coronal loops, giving rise to a soft X-ray flare.
The radio gyrosynchrotron emission (with a luminosity $L_{\mathrm{R}}$) from the accelerated 
electrons is roughly proportional to the instantaneous number of particles  and 
therefore to the power injected into the system. On the other hand, the X-ray
luminosity $L_{\mathrm{X}}$ is roughly proportional to the total energy
accumulated in the hot plasma. One thus expects, to first order,
\begin{equation} 
L_R(t) \propto {d\over dt}L_{\rm X}(t)
\end{equation}
which is known as the ``Neupert Effect'' (Neupert 1968) and has been well observed
on the Sun in most impulsive and many gradual flares (Dennis \& Zarro 1993). Only one
radio+X-ray observation of a {\it stellar} Neupert effect was obtained 
previously (G\"udel et al. 1996), additional to  one observation for which optical 
and EUV emissions were observed as the respective proxies (Hawley et al. 1995). Both 
referred to M dwarfs. RS CVn binaries are known to be surrounded by very large, 
binary-sized magnetospheres (Mutel et al. 1985) - are these coronal systems
operating the same way as a more compact solar-like corona?

\begin{figure}[t!]
\hspace{-0.2cm}
\includegraphics*[angle=0, width=6cm,height=7.9cm]{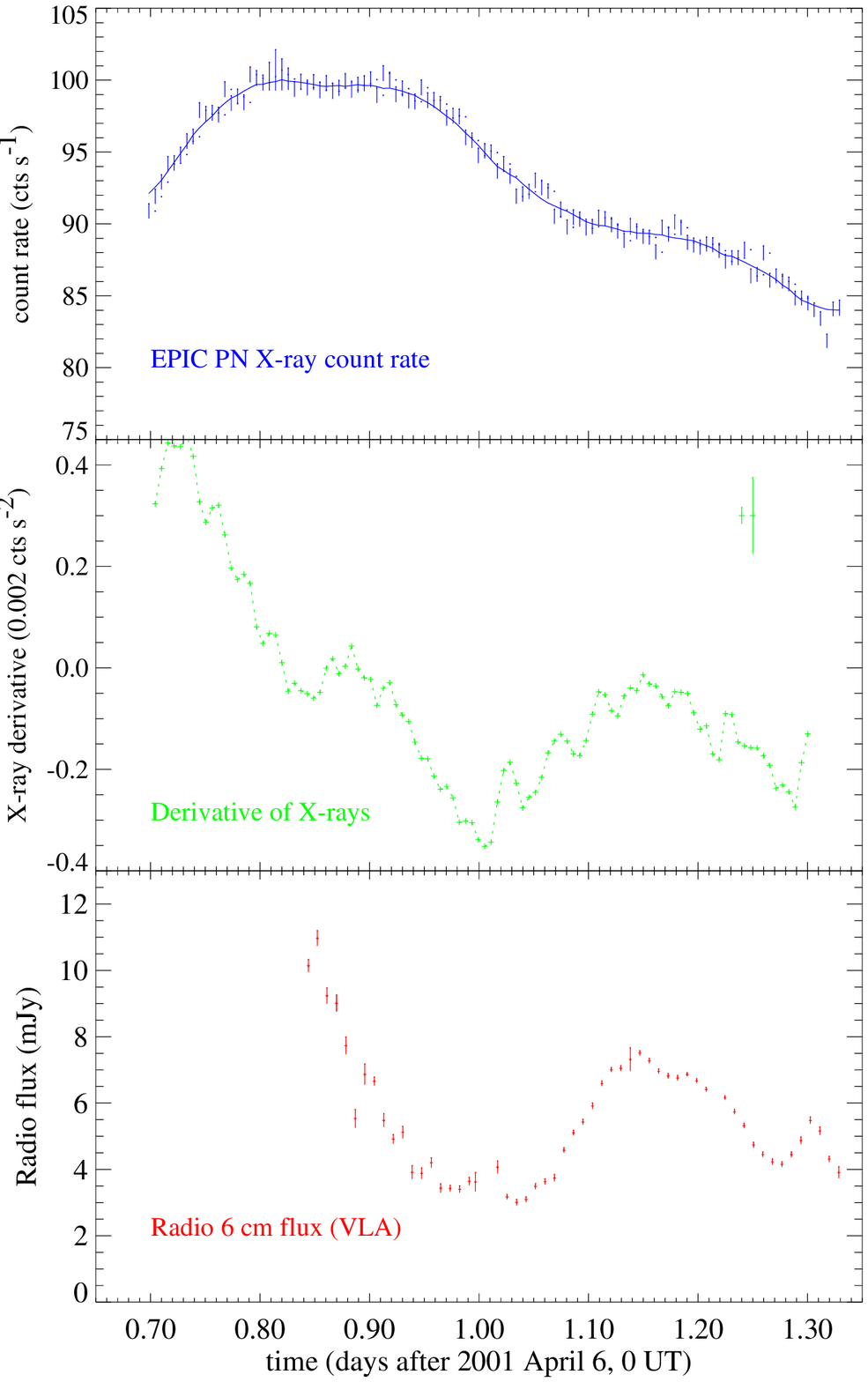}
\hspace{-0.3cm}
\includegraphics*[angle=0, width=8cm]{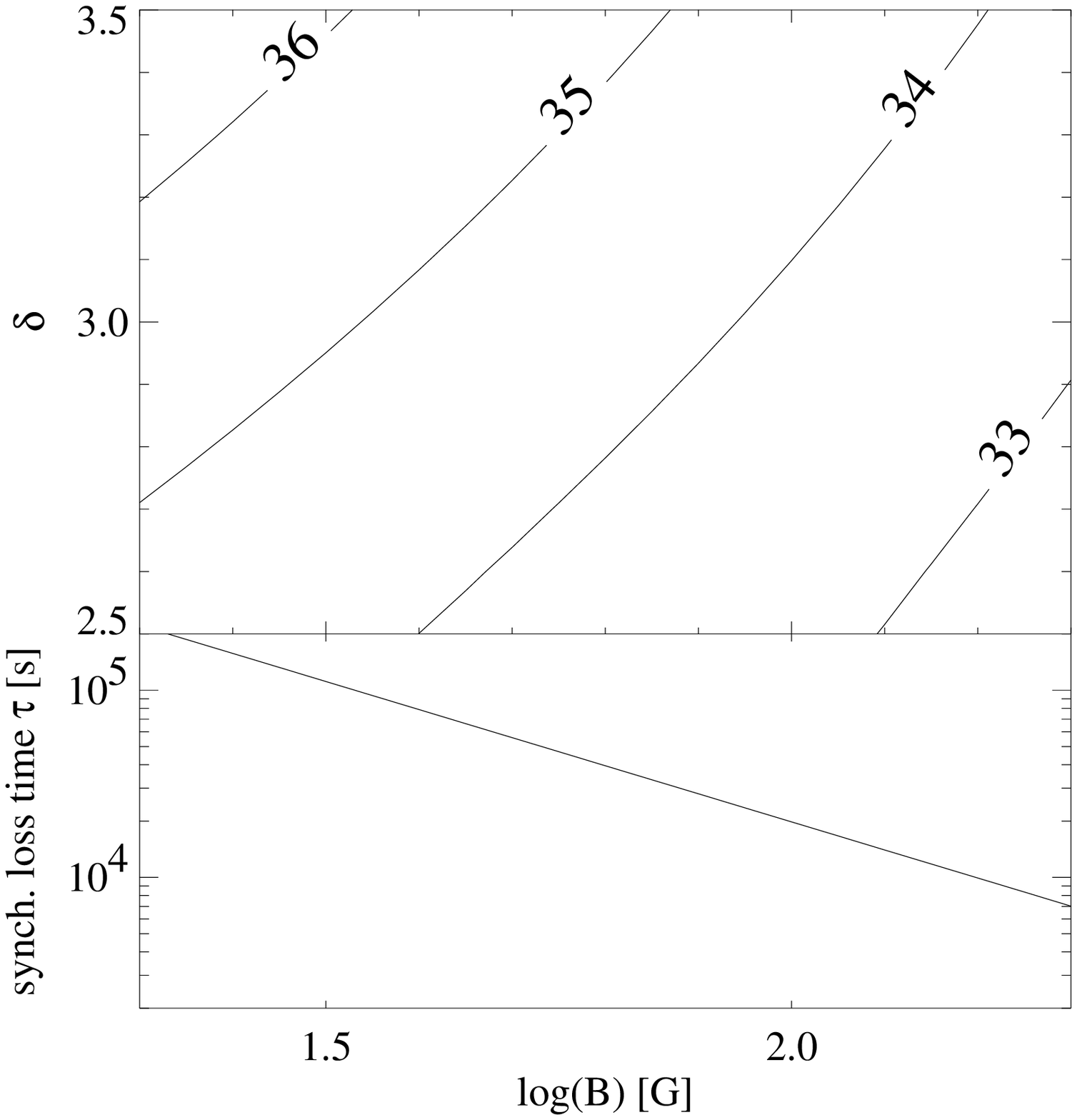}
\caption{{\bf Left:} Example of the Neupert effect in the RS CVn binary $\sigma$ Gem. 
The top panel shows the X-ray light curve, the middle panel presents the time derivative 
of the X-ray curve, and the lower panel shows the radio light curve. {\bf Right:} 
Corresponding energy content in accelerated electrons for different values of the magnetic field 
and the electron energy power-law index $\delta$ (the contour labels give the logarithms 
of the energy content in ergs). The bottom panel illustrates the synchrotron loss time for
different magnetic field strengths $B$ (after G\"udel et al. 2001c).}
\end{figure}

A recent observation with {\it XMM-Newton} and the VLA does support a solar-like
picture of energy release. Figure 3a shows, from top to bottom, the
X-ray light curve, its time derivative, and the radio light curve during a large
\index{*$\sigma$ Gem}
flare on the RS CVn binary $\sigma$ Gem. We consider only the second flare episode
that is completely covered at radio waves, i.e., the interval [1.02,1.32]~d.
The radio curve and the X-ray time derivative correlate very well
in time,  with no significant time delay. Evidently, the release of high-energy particles
is closely coupled with the heating mechanism. For chromospheric evaporation to work,
however,  a necessary condition is that the accelerated particles carry enough kinetic 
energy to explain the released soft X-ray energy. Under simplified assumptions such as an 
electron power-law distribution in energy, a reasonable lower energy cutoff around
10~keV, and the absence of strong changes in the radio optical depth, an order of
magnitude estimate of the total energy in the electrons results in a range of
$E_{\mathrm{tot}}
= 10^{33} - 10^{36}$~erg (Fig. 3b; G\"udel et al. 2001c). This estimate holds for a magnetic
field strength $B$ between 20$-$200~G and a power-law index $\delta =2.5-3.5$, values
that have previously been found from magnetospheric modeling (e.g., Mutel et  al. 1985), 
and an electron lifetime of $\sim$1500~s as inferred from the steepest parts of the
light curve (shorter lifetimes result in larger energies). The total released energy 
in the superimposed X-ray event is estimated to be $4\times 10^{34}$~erg. The kinetic particle 
energy is therefore sufficient to explain the heating for a broad range of parameters $B, 
\delta$.

\begin{figure}[t!]
\hspace{0.8cm}
\includegraphics*[angle=0, width=12cm]{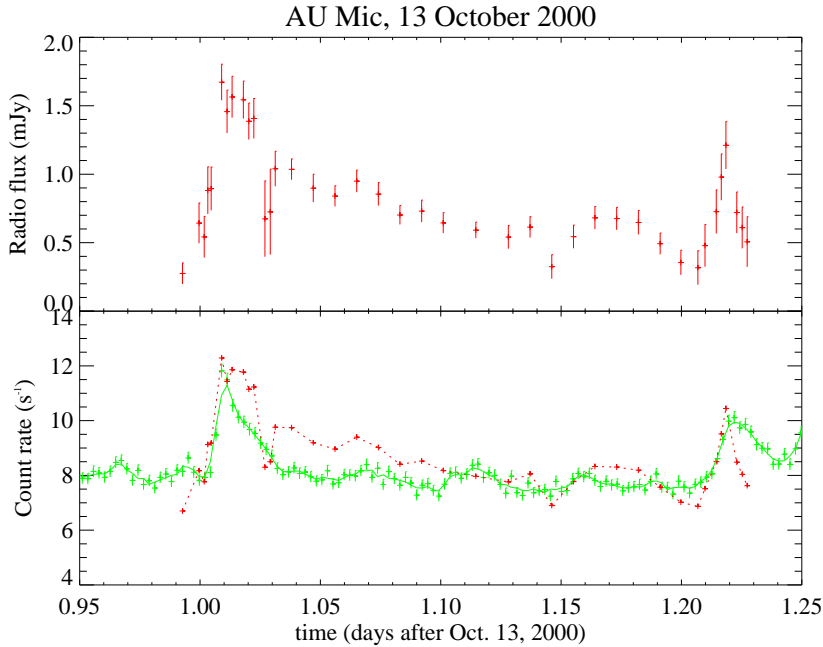}
\caption{Two flares on AU Mic, observed at radio wavelengths (top, and arbitrarily scaled 
dashed overplot at bottom) and in X-rays (bottom, error bars). The flare at 1.02~d does not show
a Neupert-effect like dependence, while the flare at 1.22~d does.}
\end{figure}

There is thus good evidence for the flare energetics in this RS CVn binary  to operate in a 
similar way as in solar flares, and for high-energy particles to play a fundamental role in
the energy release and perhaps in the plasma heating. In retrospect, we find a similar 
timing between radio and X-ray flare events in some previously published light curves, although 
the Neupert effect was not discussed. Most evidently, radio emission peaking before the soft 
X-rays, thus suggesting the presence of a Neupert effect,  can be seen in the examples presented by
Vilhu et al. (1988),  Stern et al. (1992), and Brown et al. (1998).

It is important to recognize that the Neupert effect is not a universal property of 
coronal flares, neither in the Sun (Dennis \& Zarro 1993) nor in stars. Figure 4 shows
a coordinated {\it XMM-Newton} + VLA observation of the dMe star AU Mic. While the 
second larger flare suggests a Neupert-effect dependence between the emissions, the 
first flare clearly does not. The time correlation between the two emissions is  notable,
however, testifying to concurrent particle acceleration and coronal heating.

\section{Coronal Structure - A Total Stellar X-Ray Eclipse}

\index{*$\alpha$ CrB}
\index{eclipse, total X-ray}
Eclipsing binaries are interesting for the study of coronal structure. If the eclipsing
component is dark and the eclipse is total, there is hope to  derive 
information on the location and sizes of coronal active regions on the eclipsed star.
The optically bright star $\alpha$ CrB is such a fortunate example. The geometry is sketched
in Figure 5 (upper right panel). The eclipsing star is an A star, completely dark in X-rays,
while the eclipsed object is an intermediately-active mid G-star. Eclipses occur
every $\sim$17.3 days and are total. Since the eclipse is not central, the moving limb of 
the A star cuts out differently oriented slices during ingress and egress, which should
allow for a quasi-2-D (but  non-unique) reconstruction of coronal features. The first
{\it XMM-Newton} observation on January 13, 2001, partly failed due to a guide star-acquisition 
problem so that most of the ingress  was missed. The experiment will be repeated in
summer 2001. The light curve at hand, however, already contains rather interesting 
information (Figure 5). A model assuming a homogeneous or a spherically symmetric, radially decaying
optically thin corona does not produce 
an acceptable fit to the observation (Fig. 5, top panels). We therefore calculated a
pseudo-1-D map simply from the gradients in the smoothed  light curve. The location of
the emission measure corresponding to a change in flux within a given time interval 
is unspecified within circular slices cut out by the A star (the projected corona is further 
confined to $<$1.2$R_{\mathrm{G}}$ in this example). The short segment obtained from
the ingress enhances the emission at the lower right limb, indicating that there is
an active region in that area. We thus find evidence  that the coronal
material in this star is highly structured. A full light curve will be useful to further
constrain the sizes of the active regions.

 \begin{figure}[t!]
\hspace{-0.5cm}
\includegraphics*[angle=0, width=7cm]{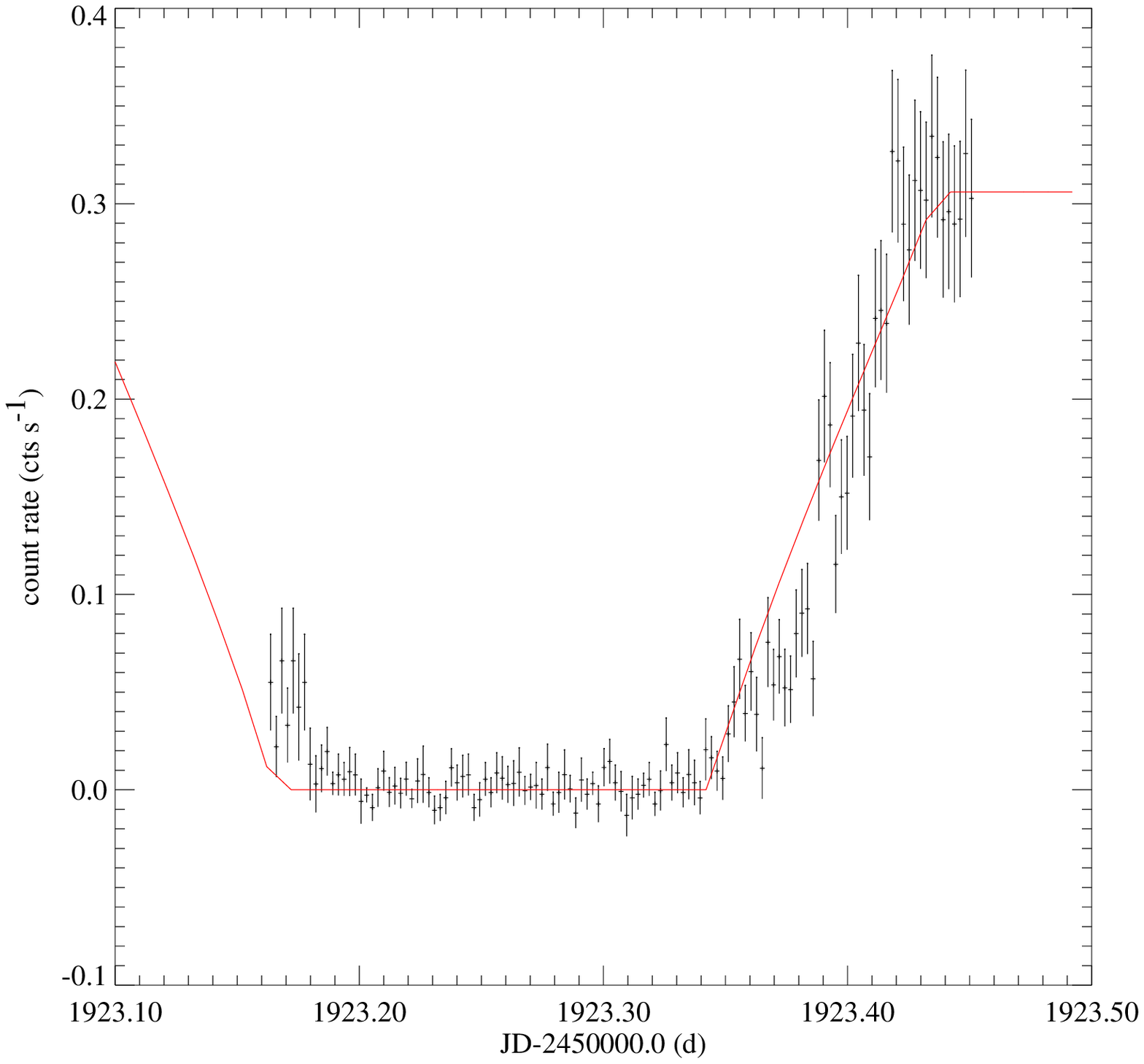}      
\hspace{1cm}
\includegraphics*[angle=0, width=6.42cm]{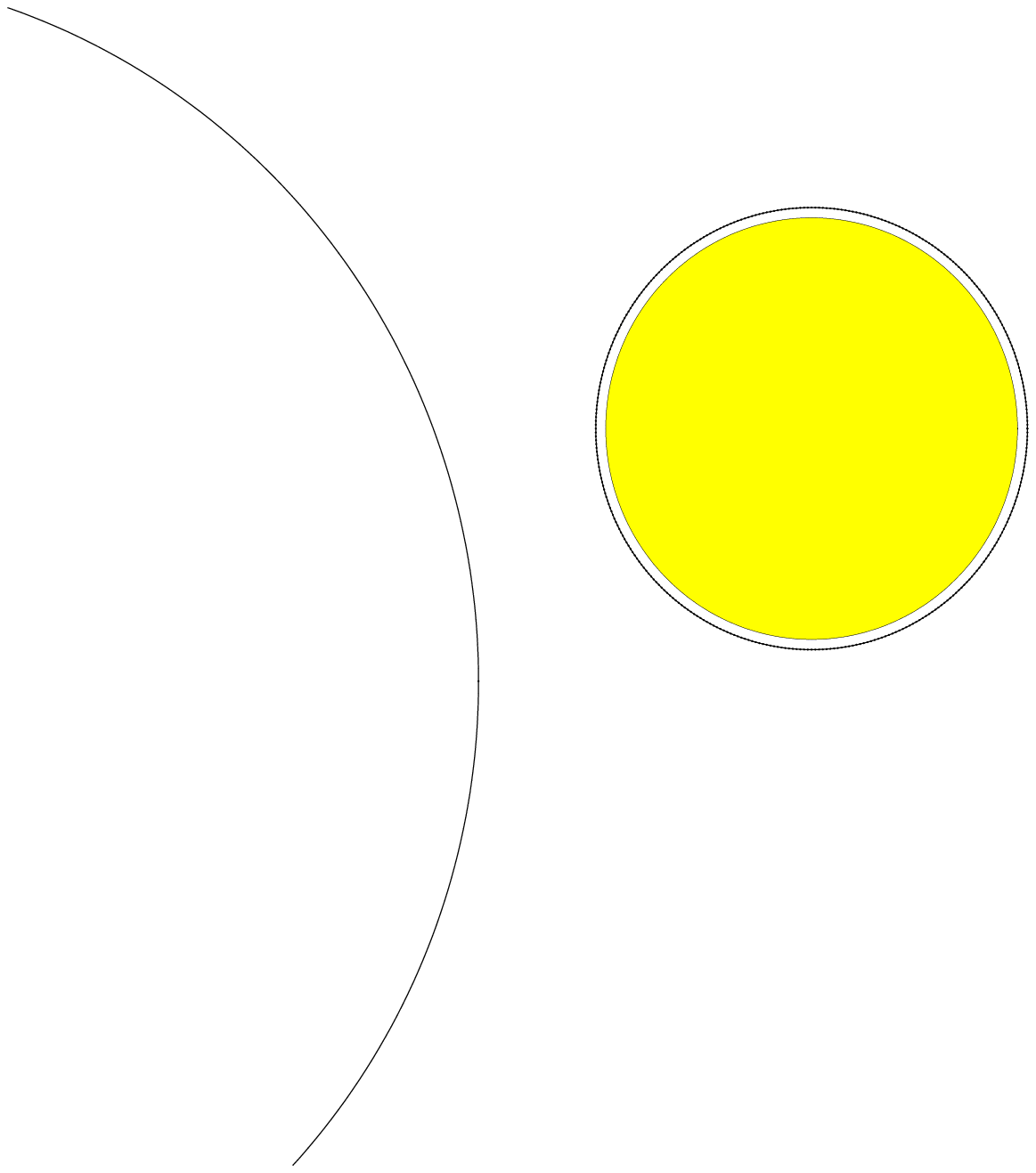}     
\hbox{\hspace{-0.4cm}\includegraphics*[angle=0, width=7cm]{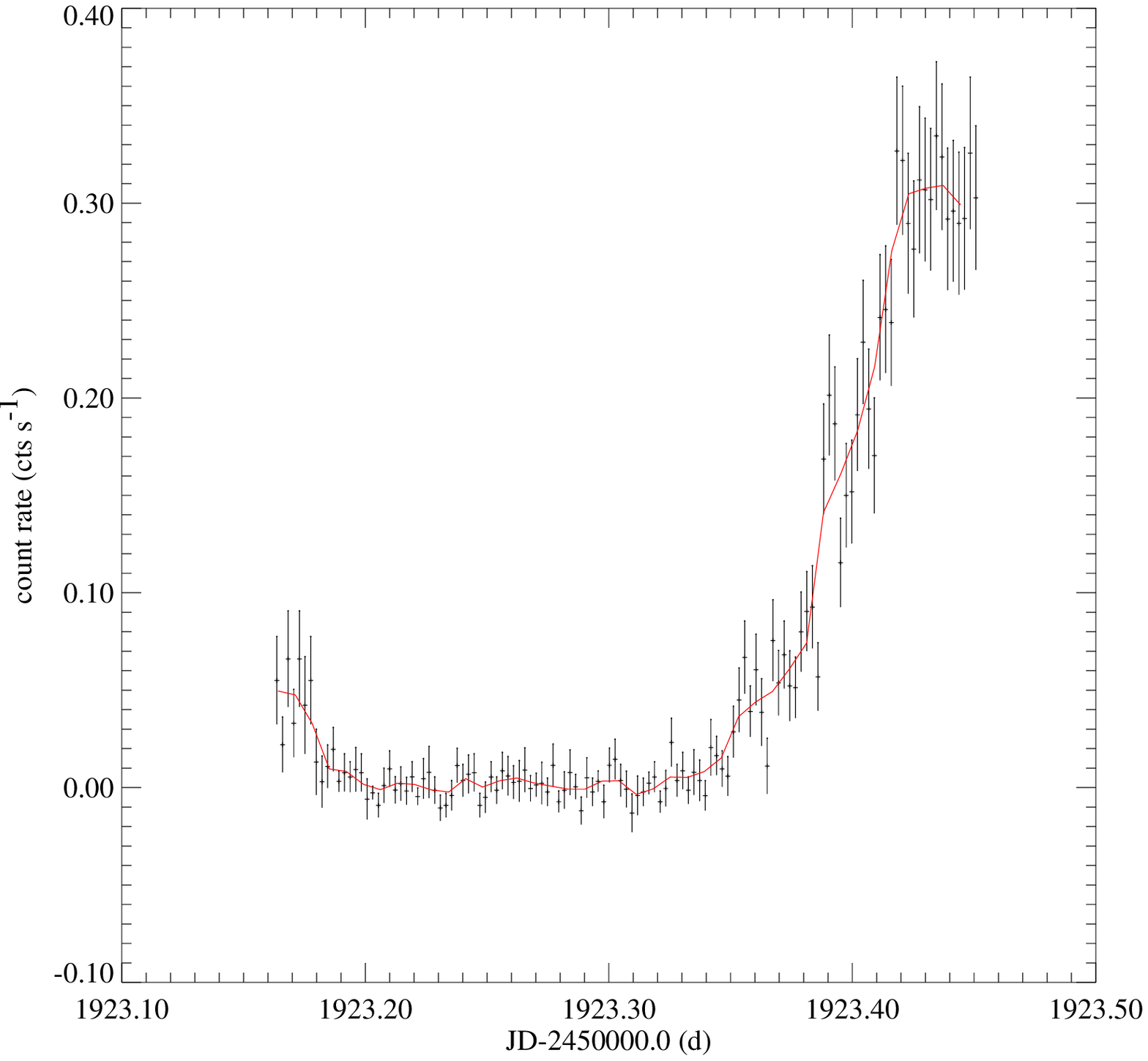}}     
\hspace{0.5cm}
\includegraphics*[angle=0, width=7cm]{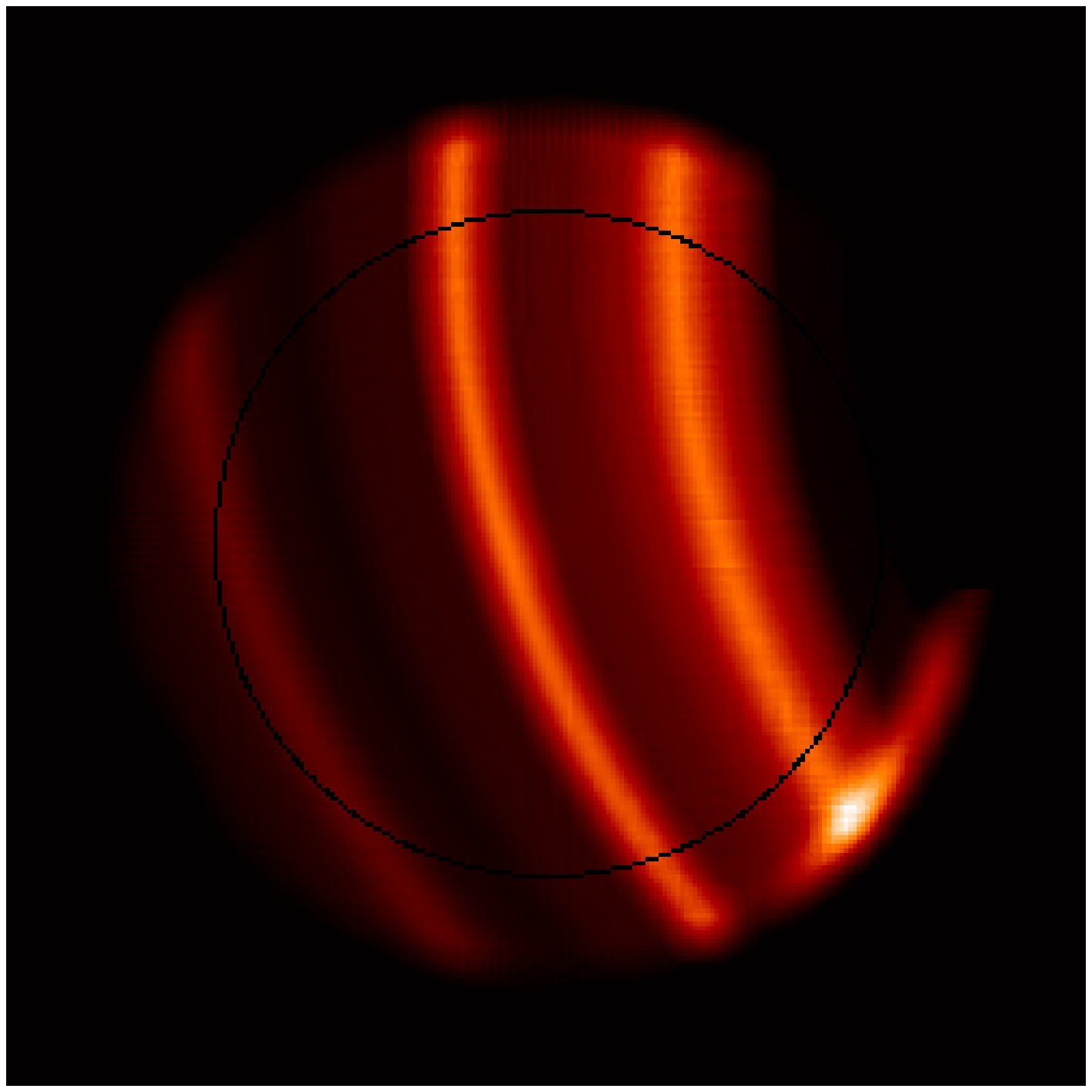}     
\caption{Total X-ray eclipse of the secondary in the $\alpha$ CrB system. {\bf Top left:}
Observed light curve and model light curve assuming a uniform, optically thin corona
extending to 1.05$R_*$, as illustrated in the {\bf top right} figure.
{\bf Bottom left:} Observed light curve and smoothed boxcar average that results in a
pseudo-1-D model shown at {\bf bottom right}.}
\end{figure}

\section{Do X-Rays Reveal Magnetic Activity Cycles?}

\index{cycle, stellar activity}
\index{*EK Dra}
We have been observing the  very active, near-ZAMS  solar analog EK Dra since 1990
using ROSAT, ASCA and EUVE. The most recent observation was obtained by {\it XMM-Newton}.
Active stars usually show irregular 
quasi-cycles of magnetic activity, although EK Dra is a clear example revealing 
a long-term behavior in magnetic spot coverage, with a cycle period around 11-12 years
as measured in optical photometry (Figure 6, top panel, Guinan et al. 2002). Does the 
activity cycle reflect
in the average X-ray luminosity? Combining snapshot data from various instruments
is not without problems and poses challenges to cross-calibration and data interpretation. 
Our best-effort result is shown in Fig. 6. A notable anticorrelation - high X-ray luminosity
during low photometric brightness or large spot coverage - is indicated. Although 
no proof for an X-ray cycle, the data are very suggestive, and a coherent continuation of 
this monitoring program would be helpful to confirm a long-term trend. 
 
\begin{figure}[t!]
\hspace{0.5cm}
\includegraphics*[angle=-90, width=12cm]{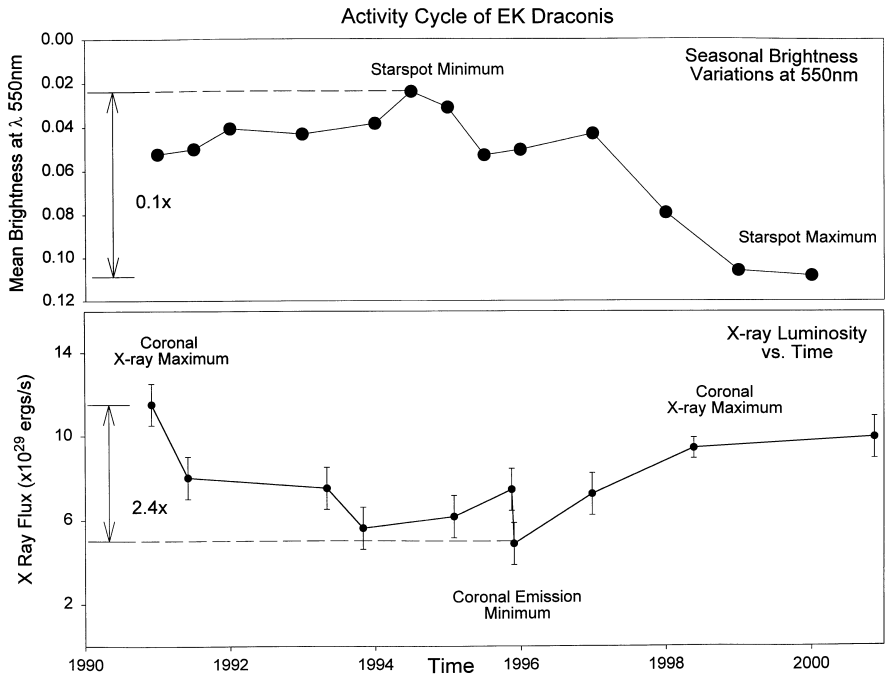}      
\caption{Photometric (top) and X-ray snapshot ``light curves'' of the young solar
analog EK Dra between 1990 and 2001. There is evidence that the starspot minimum coincides
with the lowest X-ray flux levels (Guinan et al. 2002).}
\end{figure}

\section{Summary}

A comprehensive coronal survey is underway with the instruments onboard {\it XMM-Newton}.
New standards of sensitivity and spectral resolution provide an entirely new approach
to problems in coronal physics. A summary of early results follows:
\begin{itemize}
\item There is evidence that coronal abundances follow some systematics related
to the overall coronal activity and hence the overall coronal temperature. Since
coronal activity is controlled by rotation, the coronal composition changes during main-sequence 
evolution. We speculate that high-energy processes, revealed at radio wavelengths
and indicating high-energy electrons, could be responsible for a bias in the elemental
composition of coronae of active stars.

\item We find indirect evidence for flare chromospheric evaporation on an RS CVn
binary by identifying the Neupert effect in  simultaneous radio and X-ray light curves.
Energy estimates suggest that accelerated particles play an important role in
the energy release process, and perhaps in coronal heating in general. 

\item Eclipsing binaries are ideal objects to study coronal structure. The example
of $\alpha$ CrB, a totally eclipsing system, provides clear evidence for inhomogeneity
in the G star corona.

\item There is tentative evidence for a magnetic quasi-cycle in the active, young
solar analog EK Dra both in optical photometry and in X-rays. If confirmed, it
shows that a corona near the saturation level still sensitively responds to changes 
in the photospheric activity level. A longer monitoring program is clearly warranted.
\end{itemize}

\acknowledgments
Research at PSI  has been supported by the Swiss National 
Science Foundation (grant 2100-049343). SRON is supported financially 
by NWO. The present project is based 
on observations obtained with XMM-Newton, an ESA science 
mission with instruments and contributions directly funded by 
ESA Member States and the USA (NASA). The VLA is a facility of the National Radio Astronomy 
Observatory, which is operated by Associated Universities,
Inc., under cooperative agreement with the  National Science Foundation.


\index{*Betelgeuse|see {$\alpha$ Ori}}
\index{*HD 39801|see {$\alpha$ Ori}}
\index{*HR 2061|see {$\alpha$ Ori}}
\index{*58 Ori|see {$\alpha$ Ori}}
\index{*BD +7\deg1055|see {$\alpha$ Ori}}
\index{*SAO 113271|see {$\alpha$ Ori}}
\end{document}